\documentclass[fleqn,twoside]{article}
\usepackage{espcrc2}
\usepackage{graphicx}

\title{Extragalactic Water Masers, Geometric Estimation of H$_\circ$,
and Characterization of Dark Energy} 

\author{L. J. Greenhill
   \address[]{Harvard-Smithsonian Center for Astrophysics,
              60 Garden St, Cambridge, MA 02138 USA}
    \thanks{Visiting physicist, Kavli Institute for Particle
            Astrophysics and Cosmology, Stanford University. LJG thanks
            D. Bersier, L. Macri, M. Reid, and K. Stanek for helpful discussions
            as part of ongoing collaboration. } }

\begin{document}

\begin{abstract}
High precision estimation of the equation of state of dark energy depends on
constraints external to analyses of Cosmic Microwave Background fluctuations.    A
geometric estimation of the local expansion rate, H$_\circ$, would provide the most
direct and robust constraint.  Traditional techniques to estimate H$_\circ$ have
depended on observations of standard candles for which  systematic effects can be
10\% or more.  Observations of water maser sources in the accretion disks that feed
the central engines of active galaxies enable simplified, robust, and largely
geometric analyses.  Many thousand maser sources will be discovered in studies with
the SKA, owing to its great sensitivity. Spectroscopic monitoring and interferometric
mapping -  with intercontinental baselines - will allow estimation of H$_\circ$ to
$\sim 1\%$ and possibly better.

\vspace{1pc}
\end{abstract}

\maketitle

\section{CMB Fluctuations and $H_\circ$}

Analyses of cosmic microwave background (CMB) fluctuations in the context of
power-law $\Lambda$CDM models are powerful means for estimation of cosmological
parameters, such as flatness, e.g., \cite{blake}. 
However, estimation of some parameters, such as the equation of state for dark energy,
described by $w_0$ and $\dot{w}$, requires external constraint, such as measurement
of expansion  rate.  Studies of large scale structure, lensing, and the
Sunyaev-Zeldovich effect are  being pursued with this in mind.   However, estimation
of the local Hubble constant,
$H_\circ$, directly from a sample of galaxies for which geometric estimates of
distances are available provides the most direct independent constraint (Figures 1 \&
2). Overall, estimates of H$_\circ$ with uncertainties on the order of 1\% are
desirable to resolve degeneracies in CMB analyses among $\Lambda$CDM parameters 
\cite{hu04.astroph}.

\begin{figure*}[ht]
\includegraphics[scale=0.82]{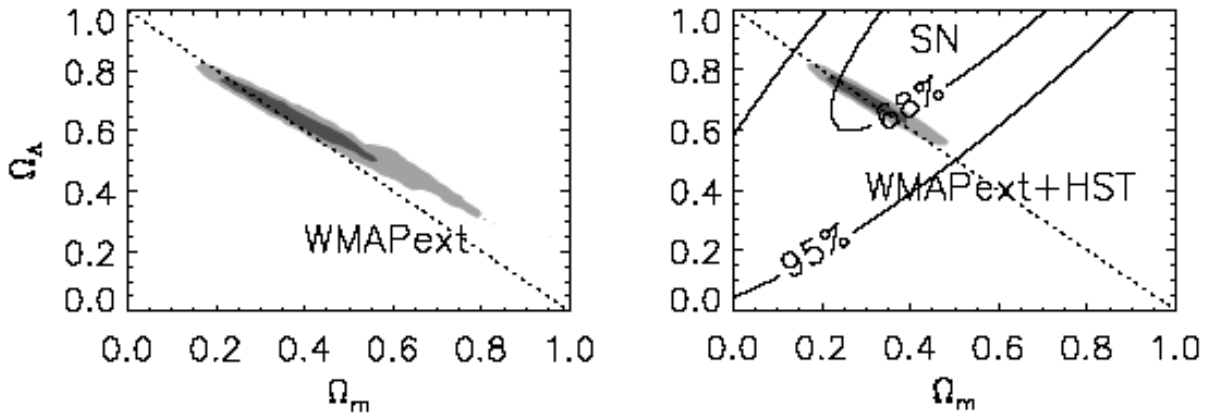}
\caption{Two-dimensional likelihood surfaces for $\Omega_m$ and
$\Omega_\Lambda$, from \cite{spergel03}, Figure~13.  {\it (left)}-- Likelihood for a
combination of several CMB experiments, the so-called ``WMAPext'' set: WMAP
\cite{bennett03}, CBI \cite{pearson03}, and ACBAR \cite{kuo04}. {\it (right)}--
WMAPext likelihood surfaces in light of independent determination of $H_\circ$
obtained by the {\it HST}\  Key Project on the Extragalactic Distance Scale, e.g.,
\cite{freedman01}, demonstrating the importance of external constraint on CMB data. 
Separate likelihoods for analysis  of  supernova data are also shown.  The dashed
line denotes a flat universe.}
\end{figure*}

\begin{figure*}[ht]
\includegraphics[scale=0.69]{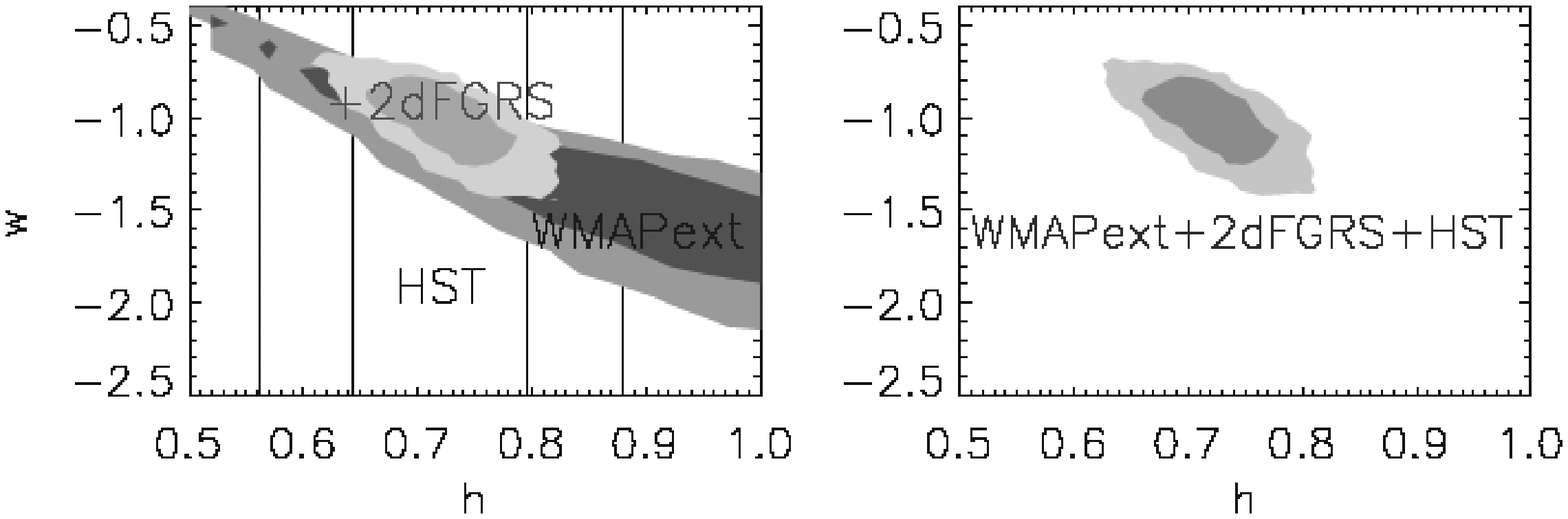}
\caption{Two-dimensional likelihood surfaces for $w$ and $h$ ($H_\circ$/100),   from
\cite{spergel03}, Figure~11.  {\it (left)}-- The outer contours are 68\% and 95\%
confidence limits obtained from for three CMB experiments shown in  Figure~1,
combined. The tighter inner contours show 68\% and 95\% limits for CMB data   combined
with data on large scale structure from the $2^\circ$-field galaxy redshift survey
\cite{colless01}.  Confidence limits for Key Project determination of the Hubble
constant, $H_\circ$, are indicated by vertical lines.  {\it (right)}-- The joint
likelihood for the equation of state of dark energy.  Key Project estimates of
H$_\circ$ have a modest impact, owing to an estimated uncertainty of 10\%.   Reduction
in that uncertainty by factors of a few would substantially improve estimates of 
$w$.}
\end{figure*}

Analysis of CMB data has been used to predict $H_\circ$ \cite{spergel03} in good
agreement with the value obtained from analysis of standard candles (i.e., Cepheid
variable stars) by the {\it HST}\/ Key Project on the Extragalactic Distance Scale
(EDS).  However, there are two limitations to this CMB analysis: the  equation of
state for dark energy ($w$) was assumed to be -1, and there is no independent check
of the  reference value of $H_\circ$ that can be used realistically to detect
deviations from $-1$. The total reported uncertainty in the reference value of
H$_\circ$ is 10\%, for analysis of 889 Cepheids in 31 galaxies \cite{freedman01}, but
control over sources of systematic uncertainty was difficult and two significant
sources stand out.

First, the best present estimate of H$_\circ$ is underpinned by the distance to the
Large Magellanic Cloud (LMC), because that galaxy is used to establish the zero point
of the Cepheid PL relation. This affects the calibration of all other standard
candles and  thus {\it the entire EDS}.  However, the suitability of the LMC as an
``anchor'' is problematic due to (1) poorly understood internal structure, e.g.,
\cite{olsen02,alves03}, and (2) controversy concerning the LMC distance,  values for
which obtained with different methods and analysis techniques disagree beyond formal
errors.  On one side a ``short'' distance modulus ($\mu_{LMC}\simeq 18.2$  mag) is
obtained with analyses of red clump stars, e.g., \cite{stanek00,udalski00} and some
detached eclipsing binaries, e.g., \cite{guinan98,fitzpatrick03}. On the other side
some estimates yield $\mu_{LMC}\simeq 18.7$, e.g., \cite{feast97}.  Confusion is
worsened because other standard candles, such as RR~Lyrae variables, support
conflicting distances, again depending on calibration and analysis method.  It is
possible the distance is different by $\pm 0.2$~mag ($\pm10$\%), twice what is
assumed in Key Project analyses.

Second, the effect of metallicity on the PL relation is controversial. Intense debate
among observers, e.g., \cite{sasselov97,kochanek97,kennicutt98,sakai04} has  not led
to broad agreement as to the sign and magnitude of the effect, while recent
theoretical work suggests sensitivity to Helium as well as metal content \cite[and
references therein]{fiorentino02,romaniello03}.  Although the Key Project adopted  the
results of \cite{kennicutt98}, with a 0.08~mag uncertainty in their
error budget \cite{freedman01}, using $VI$-band photometry alone, they  could not
disentangle the effects of reddening and metallicity. This adds uncertainty
particularly because LMC Cepheids are metal poor compared to Cepheids in our  Galaxy
and in most galaxies studied by the Key Project.  The effect of metallicity could  be
at least as large as 0.2~mag.

Conceptually, the weakest link in calibration of the EDS is that it is anchored to  a
distance  measurement for a single metal poor galaxy (the LMC), which remains 
controversial despite the application of much effort over many years.  In principal,
{\it the  EDS should be tied to a sample of ``reference'' galaxies, broadly
distributed in recessional velocity and for which robust  geometric distances are
available.  SKA mapping of water maser sources that lie in the accretion disks of
massive black holes within active galactic nuclei will make this possible. }

\section{``Geometric Anchors'' for the EDS}

For an accretion disk that is relatively edge-on, well ordered kinematically, and
heated as by X-ray irradiation or spiral shocks, conditions favor H$_2$O maser
emission (1) in a narrow sector on the near side and (2) close to the disk- diameter
perpendicular to our line of sight or ``midline,'' e.g.,
\cite{ponomarev94,miyoshi95}.  The former masers correspond to spectral features
close to the systemic velocity of the central engine.  The latter group of masers
traces the rotation curve of the disk when mapped with interferometers.  Wherever  the
emission originates, it is highly anisotropic and beamed parallel to the local  plane
of the disk.  In the case of a warped disk, the observed loci of emission mark a
compromise between where the line of sight is tangent to the disk and where  gradients
in the line-of-sight component of orbital velocities (projected along the line of
sight) is close to zero.  

\begin{figure}[ht]
\includegraphics[scale=.44]{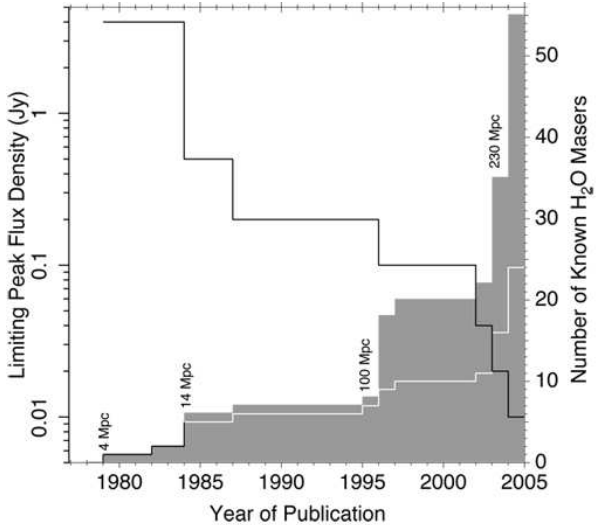}
\caption{Growth over time in the number of known megamaser sources {\it (shading)} and
the concomitant decline in peak flux density for the weakest source at each epoch {\it (black
line)}.  The increase over time in the maximum distance at which
maser emission has been recognized is also marked. The white line indicates the subset
of sources believed to be associated with edge-on accretion disks, judging from
spectroscopic signatures and/or interferometer maps.}
\end{figure} 

Two largely independent geometric estimates of distance may be obtained from
measurements of the gravitational (centripetal) acceleration and separately from
measurements of the proper motion of masers on the near side of the disk.  The
acceleration is obtained from the secular change of maser line-of-sight velocities
(observed in spectra), and the proper motion is obtained from the changes in  angular
position (observed in interferometer images).  Successful estimation of distance
depends on a disk exhibiting emission from the near side of the disk, to enable
measurement of acceleration or proper motion, and emission from the midline, to
enable modeling of the disk geometry (e.g., inclination) and central mass.

To date, a geometric distance has been estimated for one  galaxy, NGC\,4258, and  the
example is a pathfinder for future work involving other galaxies.  Herrnstein et  al.
\cite{herrnstein99} reported acceleration and proper motion distances that agree  to
$<1\%$, and they obtained random and systematic uncertainties of  4\% and 6\%,
respectively.  The systematic uncertainty arises largely from unmodeled  substructure
and a 0.1 upper limit on the eccentricity of orbits in the 0.56 pc diameter disk.   An
increase in the number of observing epochs from 6 to 24, extension of the time
baseline of monitoring from 3 to 6 years, and incorporation of eccentricity into  the
geometric model may ultimately reduce the total uncertainty in the geometric  distance
to $\sim 3\%$ \cite{humphreys04}.

However, using the distance to NGC\,4258 to estimate $H_\circ$ is complicated  because
the galaxy is nearby and its peculiar motion above and beyond the Hubble expansion  is
large. A geometric distance for NGC\,4258 is most appropriately used to  recalibrate
Cepheids PL relations independent of the LMC distance and sub-solar metallicity.   As
a result, uncertainty in a new estimate of H$_\circ$ would combine error budgets for
maser and Cepheid analyses.  Ultimately, the best results will be achieved for
galaxies distant enough that their peculiar motions are a small fraction of their
total motions, and H$_\circ$ may be estimated directly from maser distances and
recessional velocities.

\section{SKA Measurement of $H_\circ$}

The sensitivity of the SKA will be critical to assembling a large sample of
maser-host galaxies and estimating their distances.  Somewhat more than 50 masers are known at flux
densities above $\sim 10$~mJy and distances of $\sim 4$ to 230~Mpc.  This count
is strongly sensitivity limited.  Larger and more efficient apertures detect more
masers, and rapid growth in the number of known masers since 2000 is primarily a
consequence of improving instrument sensitivity (Figure~3).  The best detection  rates
achieved thus far are for the Green Bank Telescope (GBT),  $\sim 30\%$ for Seyfert-2 
galaxies with $cz<5000$~km\,s$^{-1}$, of which there are 34 masers
\cite{braatz04,kondratko04}.  The SKA will be $\sim 80\times$ more sensitive than  the
GBT at $\lambda1.3$cm.  Judging from the larger volume of space that the SKA will
explore with at least comparable sensitivity, it may be expected to increase  the
number of known H$_2$O masers by two to three orders of magnitude \cite{morganti}.

Estimating  distances for galaxies beyond a few tens of megaparsecs is best
accomplished by measurement of centripetal accelerations through monitoring of
spectra, and by measurement of angular structure via interferometric mapping.  For  a
zero order model  disk (i.e., flat, effectively massless, and otherwise similar to
NGC\,4258), $D = a^{-1}(\theta_h V_{rot}^2|_{\theta_h})^{1\over3} 
\Omega^{4\over3}$,   where $D$ is distance, $a$ is centripetal acceleration,
$V_{rot}|_{\theta_h}$ is rotation speed at angular radius $\theta_h$ (estimated from
a measured rotation curve),  and $\Omega$ is orbit curvature estimated for material
along the near side of the disk (obtained to first order from position-velocity plots
of interferometer data).

The overall factional uncertainty for an individual distance
measurement, $\sigma_D\over D$,  is approximately

\begin{equation}
{1\over3}\sqrt{ 4({ \sigma_{V_{rot}} \over
V_{rot}})^2 + ({\sigma_{\theta_h} \over \theta_h})^2 + 9({\sigma_a\over a})^2
+ 16({\sigma_\Omega \over \Omega})^2 }
\end{equation}

\noindent
where $\sigma$ is used to indicate measurement uncertainty.  Consider a nominal
10~mJy maser source that is observed with the SKA for one hour to construct images
with sub-milliarcsecond resolution and whose spectrum is monitored with one minute
snapshots every two months for one year.  A spectral line width of 3~km\,s$^{-1}$,
which is typical, is assumed for  the following calculations.  Adopting a sensitivity
of $\sim 1$~mJy\,hr$^{-0.5}$ for very long baseline imaging (i.e., for the core of
the SKA operating in tandem with outrigger antennas that provide intercontinental
baselines) \cite{morganti}  and a sensitivity for spectroscopy corresponding to
$A_{\rm e}/T_{\rm sys}\sim 10^4$~m$^2$\,K$^{-1}$ (at $\lambda1.3$~cm), one obtains
${\sigma_D\over D}< 10\%$ for a wide range of  central engine mass, disk radius, and
distance (Figure~4).  In principle, actual  measurements may be affected by
systematics, as may be introduced by blending of spectral lines, nonKeperlian rotation 
curves or substructure within the disk (e.g., the Circinus maser; \cite{greenhill03}), but
it is difficult to assess their magnitude in advance.  If the case of NGC4258  is typical
of the subsample that will be studied in detail to obtain H$_\odot$, then the systematic
and random errors will be the same order of magnitude, and this may suggest an upper
limit of $<20\%$ on $\sigma_D/D$ for individual galaxies.  

As discussed by Morganti et al., the SKA is expected to detect many thousand water
maser sources in the accretion disks of massive black holes.  Conservatively, on the
order of 10\% will be useful for the estimation of distances, based on expectations
developed in light of the source sample known today.  (The actual fraction could 
turn out to be substantially larger.)  The
systematic uncertainties that affect individual distance measurements will be largely
uncorrelated.  Overall uncertainty in  H$_\circ$ will scale inversely with
$\sqrt{N}$, where $N$ is the number of distance measurements (assuming peculiar
motions are small), and a 1\% factional uncertainty on H$_\circ$ will
be achievable with no more than just a few hundred ``maser galaxies,'' an order
of magnitude improvement over current best estimates obtained by the study of
standard candles.  The impact on parameter estimation for
$\Lambda$CDM or other precision cosmological models will be substantial (e.g.,
\cite{hu04.astroph}) and would be difficult to obtain by other means, because the
maser studies will rely primarily on geometry and the results will be largely model
independent.  

The critical elements that will enable high accuracy measurement of H$_\circ$ are
(1)~a short wavelength capable SKA  ($\lambda1.3$~cm), (2)~outrigger antennas that
contribute substantial collecting area on  intercontinental baselines ($A_{\rm
e}/T_{\rm sys}$ on the order of $10^3$~m$^2$\,K$^{-1}$), (3)~high spectral
resolutions and broad instantaneous bandwidths (${\Delta\nu\over\nu}\sim10^{-6}$) to
resolve lines distributed over of order 2000~km\,s$^{-1}$, (4)~time to observe tens
of thousands of active galactic nuclei with the goal of identifying new maser sources
and (5)~follow-up imaging and spectroscopic monitoring of the hundred or thousands 
of new maser sources, keeping in mind that long synthesis tracks will be required to
achieve adequate sensitivity with the highest angular resolutions.

\begin{figure*}[ht]
\includegraphics[scale=0.95]{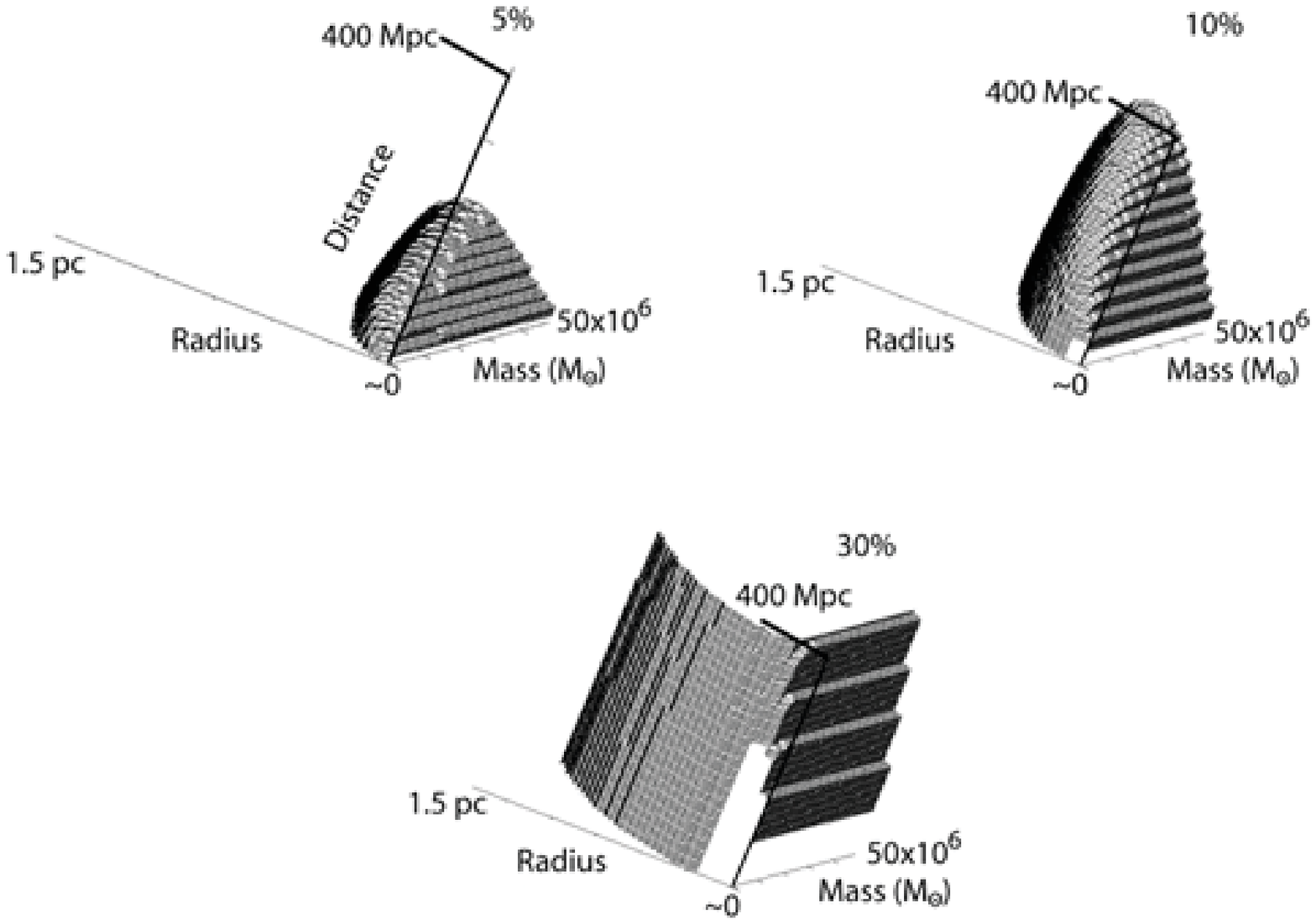}
\caption{Surfaces that represent fractional uncertainties in estimated distance of 5\%,
10\%, and 30\% (for an individual galaxy) as functions of central engine mass, accretion
disk radius, and true distance. The effects of systematic uncertainties are not included,
but in the case of NGC\,4258, random and systematic uncertainties in estimated distance
are about the same magnitude.  }

\end{figure*}

\end{document}